\documentclass[aps,prb,bibnotes,twocolumn,showpacs,preprintnumbers,amsmath,amssymb,superscriptaddress,floatfix]{revtex4-2}
\usepackage[T1]{fontenc}
\usepackage[ansinew]{inputenc}
\usepackage{graphics}
\usepackage{dcolumn}
\usepackage{bm}
\usepackage{amsthm}
\usepackage{amsmath}
\usepackage{amssymb}
\usepackage{diagbox}
\usepackage{hyperref}
\usepackage{url}
\usepackage{xcolor}
\usepackage{soul}

\begin{document}

\title{Angular anisotropy landscape of vortex ensembles in polarized small-angle neutron scattering}

\author{Michael P.\ Adams}
\affiliation{Max Planck Institute for the Structure and Dynamics of Matter, Luruper Ch.~149, D-22761 Hamburg, Germany}
\author{Elizabeth M.\ Jefremovas}
\affiliation{Department of Physics and Materials Science, University of Luxembourg, 162A Avenue de la Faiencerie, L-1511 Luxembourg, Grand Duchy of Luxembourg}
\affiliation{Institute for Advanced Studies, University of Luxembourg, L-4365 Esch-sur-Alzette, Luxembourg}
\author{Andreas Michels}
\affiliation{Department of Physics and Materials Science, University of Luxembourg, 162A Avenue de la Faiencerie, L-1511 Luxembourg, Grand Duchy of Luxembourg}

\begin{abstract}
We present a symmetry-resolved classification of two-dimensional spin-flip small-angle neutron scattering (SANS) patterns arising from dilute ensembles of spherical nanoparticles hosting magnetic vortex states. Based on a linear vortex ansatz with an axially symmetric distribution of vortex axes and the corresponding analytical expression for the orientationally averaged spin-flip SANS cross section, we show that the angular scattering patterns organize into four distinct symmetry regimes: a four-fold anisotropy corresponding to coherent field-aligned magnetization, vertical and horizontal two-fold anisotropies associated with aligned and isotropically distributed vortex ensembles, and an isotropic ring-like condition separating the two two-fold regimes. The corresponding symmetry boundaries are obtained analytically and define a compact symmetry landscape in the parameter space of vortex amplitude and vortex-axis distribution width. Comparison with a nonlinear vortex profile shows that these symmetry regions are robust with respect to the detailed radial structure of the vortex core. The angular anisotropies are therefore governed primarily by rotational symmetry and by the statistical distribution of vortex axes, providing a compact and model-transparent classification framework of experimental polarized SANS data.
\end{abstract}


\maketitle


\section{Introduction}

Magnetic small-angle neutron scattering (SANS) has become a key technique for probing nanoscale spin textures in bulk materials and nanoparticle assemblies (see, e.g., Refs.~\cite{bersweiler2019size,zakutna2020,laura2020,dirkreview2022,gerina2023size,bersweilerprb2023,titov2025,borchers2025,disch2025} and references therein). In polarized SANS experiments, the two-dimensional spin-flip scattering cross section encodes the angular anisotropy of the purely magnetic correlations in reciprocal space. Depending on magnetic field and material composition, characteristic two-fold and four-fold intensity distributions have been observed experimentally~\cite{bender2018dipolar,benderapl2019}. However, relating these angular symmetries to the underlying real-space spin textures remains nontrivial, particularly in systems with noncollinear magnetization and collective dipolar interactions.

Vortex-like magnetization states constitute a prototypical example of such textures (e.g., \cite{Bonilla2017,Usov2018,Niraula2023_VortexNanospheres,BATLLE2024,jefremovas2026coercivity,Adams2026minimal}). These are predicted to emerge in confined ferromagnetic structures that exceed the single-domain size limit as a result of the competition between exchange, magnetostatic, and anisotropy energies~\cite{aharonibook}.

A linear vortex model was recently introduced to describe the magnetic spin-flip SANS cross section of dipolar-stabilized nanoparticle ensembles and was validated against micromagnetic simulations~\cite{Adams2024vortex}. While this framework provides an analytical description of the scattering cross section, its full angular solution space has not been investigated in a symmetry-resolved manner. In particular, it remains unclear how the interplay between vortex amplitude and vortex-axis disorder governs the symmetry class of the observable scattering pattern.

In the present work, we analyze the complete angular solution space of the linear vortex ensemble model. Starting from the closed-form analytical expression of the orientationally averaged spin-flip SANS cross section~\cite{Adams2024vortex}, we systematically explore the two-dimensional parameter space defined by vortex amplitude and the angular width of the vortex-axis distribution. We demonstrate that the possible scattering patterns collapse into four distinct symmetry regimes: four-fold, vertical two-fold, horizontal two-fold, and isotropic ring-like.

The resulting symmetry landscape is represented as a classification map of angular anisotropies for magnetic vortex-dominated spherical nanoparticle systems. By linking experimentally observed scattering patterns to regions in parameter space, our work establishes a compact and model-transparent classification framework for polarized SANS data. All results in this work refer to the scattering geometry where the applied magnetic field $\mathbf{H}_0$ is perpendicular to the wave vector $\mathbf{k}_0$ of the incoming neutron beam ($\mathbf{H}_0 \perp \mathbf{k}_0$)~\cite{Adams2024vortex}.


\section{Linear Vortex Ensemble Approach}
\label{VortexTheory}

\begin{figure}
    \centering
    \resizebox{1.0\columnwidth}{!}{\includegraphics{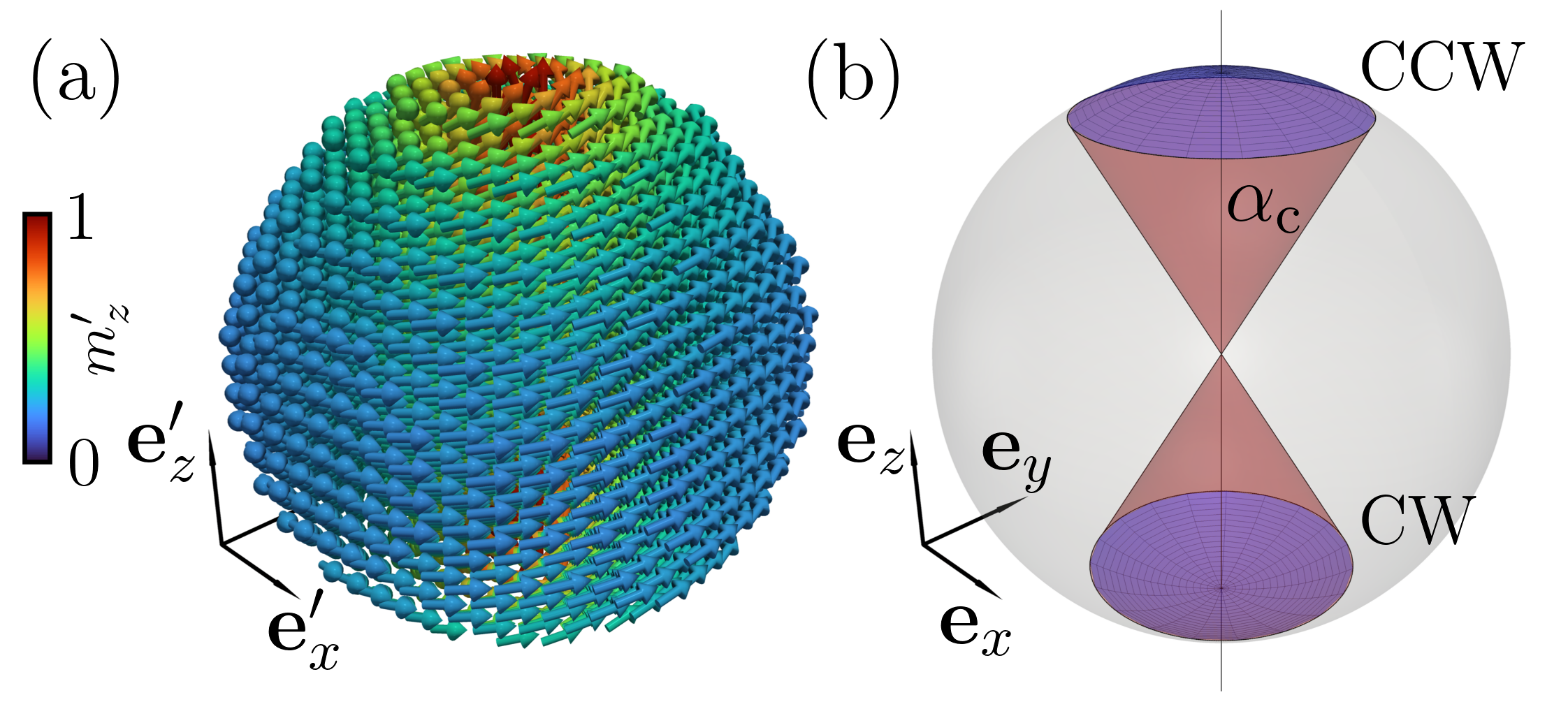}}
    \caption{
    (a)~Vortex configuration inside a spherical particle. The color scale encodes the longitudinal magnetization component $m'_z$. (b)~Representation of the vortex-axis ensemble on the unit sphere. The red double cone defines the angular cutoff angle $\alpha_{\mathrm{c}}$ with respect to the $\mathbf{e}_z$~axis of the global laboratory frame. The blue spherical caps correspond to the allowed statistical distribution of vortex axes. CW and CCW vortices populate the upper and lower caps with equal probability, reflecting the absence of chirality selection.
    }
    \label{fig1}
\end{figure}

Dipolar-dominated nanoparticle assemblies naturally develop vortex-like magnetization textures [see Fig.~\ref{fig1}(a)] in order to minimize the magnetostatic energy~\cite{aharonibook}. In the absence of inversion-symmetry-breaking interactions (e.g., the Dzyaloshinskii-Moriya interaction), no energetic chirality selection is expected. Consequently, clockwise (CW) and counterclockwise (CCW) vortices occur with equal probability [Fig.~\ref{fig1}(b)].

The lowest-order continuum description of such a vortex state is provided by the following linear vortex ansatz. In the local particle coordinate frame $(x', y', z')$, we write
\begin{equation}
\mathbf{m}'(\mathbf{r}')
=
\left[ m_{0} \, \mathbf{e}_{z}'
+
p \, m_{1} \, \rho' \, \mathbf{e}_{\varphi}' \right] \, \Theta(1 - r'/R) ,\label{eq:LinearVortexAnsatz}
\end{equation}
where $m_{0}$ denotes the amplitude of the uniform axial component, $m_{1}$ controls the amplitude of the curling in-plane component,
$\mathbf{e}_z' = \{0, \, 0,  \, 1\}$ and $\mathbf{e}_{\varphi}' =\{-\sin\varphi', \, \cos\varphi', \, 0\}$ are the cylindrical unit vectors, $\rho' = \sqrt{(x')^2 + (y')^2}$ and $\varphi' = \arctan(y'/x')$ are the cylindrical coordinates, $\Theta$ is the Heaviside step function, $r'=\sqrt{(\rho')^2 + (z')^2}$ is the spherical radial coordinate, $R$ is the particle radius, and $p \in \{-1,1\}$ specifies the chirality. A chirality of $p=1$ corresponds to CCW and $p=-1$ to CW curling. Despite its simplicity, this ansatz captures the dominant real-space and reciprocal-space signatures of vortex states (see Ref.~\cite{Adams2024vortex} for details).

In realistic nanoparticle ensembles, however, not only the chirality but also the vortex-axis orientation varies throughout the ensemble. The magnetization field in the laboratory frame, $\mathbf{m}(\mathbf{r})$, is therefore obtained by rotating the local vortex configuration $\mathbf{m}'(\mathbf{r'})$, according to:
\begin{equation}
\label{eq:LinearVortexModeltransformed}
\mathbf{m}(\mathbf{r})
=
\mathbf{R}(\alpha,\beta) \cdot
\mathbf{m}'\!\left(
\mathbf{R}^T(\alpha,\beta) \, \mathbf{r}
\right),
\end{equation}
where $\mathbf{R}(\alpha,\beta)$ denotes the rotation matrix that aligns the local vortex axis with the unit vector
\begin{align}
\mathbf{u}(\alpha, \beta)
&=
\frac{\nabla \times \mathbf{m}}{|\nabla \times \mathbf{m}|} 
\\
&=
p \sin\alpha \cos\beta \, \mathbf{e}_x
+
p \sin\alpha \sin\beta \, \mathbf{e}_y
+
p \cos\alpha \, \mathbf{e}_z. \nonumber
\end{align}
This construction defines not a single vortex, but an \emph{ensemble of vortices} characterized by a statistical
distribution of axis orientations on the unit sphere.

We model the orientational distribution of vortex axes by an axially symmetric cone with cutoff angle $\alpha_{\mathrm{c}}$,
\begin{align}
\psi(\alpha, \beta)
=
\frac{\Theta(\alpha_{\mathrm{c}} - \alpha)}
{2\pi (1 - \cos\alpha_{\mathrm{c}})} ,
\label{eq:UniformDistribution}
\end{align}
where $\int_0^{2\pi} \int_0^{\alpha_{\mathrm{c}}} \psi(\alpha, \beta) \sin\alpha d\alpha d\beta = 1$. The cutoff angle $\alpha_{\mathrm{c}}$ parametrizes the degree of vortex-axis alignment, which experimentally can be achieved by tuning the applied magnetic field strength.

To connect the mesoscale vortex ensemble to experimentally accessible observables, we employ the multinanoparticle power-series expansion (MNPSE) method~\cite{Adams2024vortex,adams2024signature}. In the dilute limit (no interparticle interference scattering), the ensemble-averaged spin-flip SANS cross section reads
\begin{equation}
\left\langle \frac{d\Sigma_{\mathrm{sf}}}{d\Omega} \right\rangle
=
\frac{1}{2}
\int
\left[
\frac{d\Sigma_{\mathrm{sf}}^{\mathrm{CCW}}}{d\Omega}
+
\frac{d\Sigma_{\mathrm{sf}}^{\mathrm{CW}}}{d\Omega}
\right]
\psi(\alpha,\beta) \, d\Upsilon,
\end{equation}
where $d\Upsilon = \sin\alpha d\alpha d \beta$. Carrying out the orientational averaging yields the analytical two-dimensional spin-flip SANS cross section
\begin{widetext}
\begin{align}
\label{eq:SpinFlip2DSANScrossSection}
\left\langle\frac{d\Sigma_{\mathrm{sf}}}{d\Omega}\right\rangle(q,\theta)
&\propto
\frac{1}{4} \, [m_0 f(qR)]^2
\Big[
12
- (\cos^2\alpha_{\mathrm{c}} + \cos\alpha_{\mathrm{c}})
(3\cos^2(2\theta) + 2\cos(2\theta) + 3)
+ 4\cos(2\theta)
\Big] \\
&\quad
+ [m_1 R f'(qR)]^2
\Big[
3
- (2\cos^2\alpha_{\mathrm{c}} + 2\cos\alpha_{\mathrm{c}} - 1) \cos(2\theta)
\Big] \nonumber ,
\end{align}
\end{widetext}
with $f(u) = (\sin u - u\cos u)/u^3$ and $f'(u) = df(u)/du$. Equation~\eqref{eq:SpinFlip2DSANScrossSection}
forms the basis of the symmetry landscape analysis presented below.

\begin{figure*}
    \centering
    \resizebox{0.95\textwidth}{!}{\includegraphics{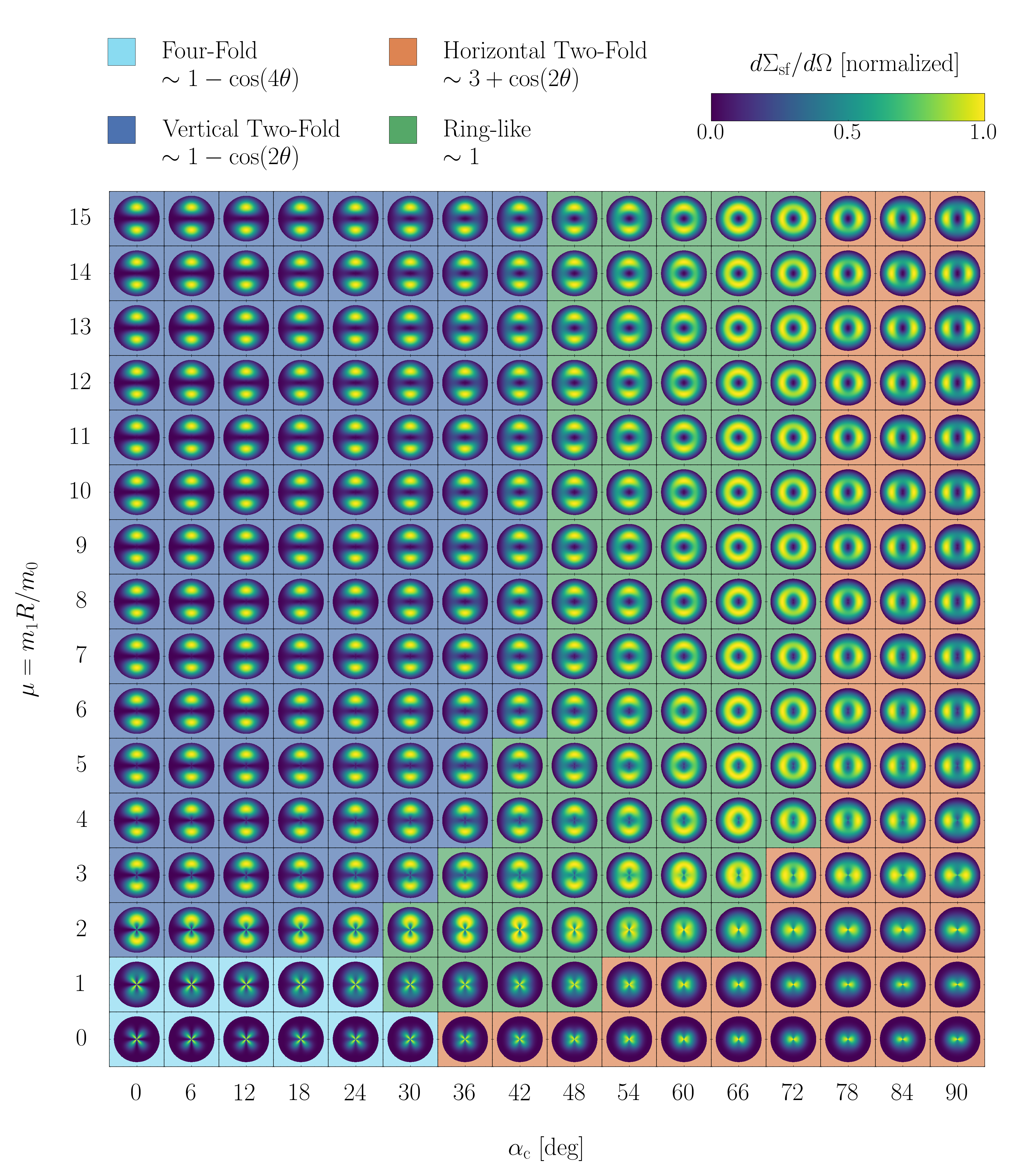}}
    \caption{
        Symmetry-resolved map of the (normalized) two-dimensional spin-flip SANS cross section for a linear vortex ensemble. Each panel shows the analytically calculated spin-flip scattering pattern [up to $(qR)_{\mathrm{max}} = 5.0$] obtained from Eq.~(\ref{eq:SpinFlip2DSANScrossSection}) for a given combination of the vortex amplitude parameter $m_1 R / m_0$ (vertical axis) and the cone angle $\alpha_{\mathrm{c}}$ (horizontal axis), which describes the statistical distribution of vortex axes within a rotational cone. The model represents an ensemble of linearly varying vortex structures whose axes are distributed within a cone of opening angle $\alpha_{\mathrm{c}}$, rather than a single vortex configuration. The background colors indicate the dominant angular anisotropy of the scattering pattern: four-fold symmetry $\propto 1 - \cos(4\theta)$ (light blue), vertical two-fold symmetry $\propto 1 - \cos(2\theta)$ (dark blue), horizontal two-fold symmetry $\propto 3 + \cos(2\theta)$ (orange), isotropic ring-like scattering $\propto 1$ (green), and intermediate transition regions. The map provides a controlled visual classification tool for experimental two-dimensional spin-flip SANS data, linking observed angular anisotropies to specific regions in the parameter space of vortex amplitude and axis distribution.
        }
    \label{fig2}
\end{figure*}

\section{Landscape of angular anisotropy patterns}

Figure~\ref{fig2} presents the classification map of the two-dimensional spin-flip SANS patterns obtained from Eq.~\eqref{eq:SpinFlip2DSANScrossSection} as a function of the vortex amplitude $m_1$, the axial amplitude $m_0$, and the cone angle $\alpha_{\mathrm{c}}$. Introducing the dimensionless ratio
\[
\mu = \frac{m_1 R}{m_0}
\]
reduces the parameter space to the two variables $(\mu,\alpha_{\mathrm{c}})$. For each parameter pair, the calculated spin-flip cross section was normalized to its maximum intensity in order to emphasize the angular symmetry, independent of the absolute scattering intensity.

The patterns in Fig.~\ref{fig2} are grouped into four angular anisotropy classes: four-fold, vertical two-fold, horizontal two-fold, and ring-like. The classification is based on the radially integrated angular dependence
\begin{align}
I(\theta)
=
\int_{q_{\min}}^{q_{\max}}
\left\langle \frac{d\Sigma_{\mathrm{sf}}}{d\Omega}\right\rangle(q, \theta) \, dq ,
\end{align}
which for Eq.~\eqref{eq:SpinFlip2DSANScrossSection} leads to the form
\begin{align}
I(\theta)
=
A(\alpha_{\mathrm{c}})
+
B(\alpha_{\mathrm{c}}) \cos(2\theta)
+
C(\alpha_{\mathrm{c}}) \cos(4\theta) ,
\end{align}
where the $A, B, C$ are functions of $\alpha_{\mathrm{c}}$. The angular structure is therefore fully determined by the relative weight of the $\cos(2\theta)$ and $\cos(4\theta)$ contributions. For each $(\mu,\alpha_{\mathrm{c}})$, $I(\theta)$ is classified by least-squares comparison with the four analytically derived limiting forms of the model discussed below, corresponding to the four-fold, vertical two-fold, horizontal two-fold, and ring-like reference states. The class was assigned according to the smallest root-mean-square deviation. This discrete assignment is adopted intentionally in order to provide a compact symmetry-based classification of the radially integrated patterns.

\textit{(i)~Saturated state---four-fold ($\mu = 0, \alpha_{\mathrm{c}} = 0$).}
The fully saturated ensemble under an infinitely strong applied magnetic field is characterized by a uniform magnetization in each particle ($\mu = 0$) and a perfectly aligned orientational distribution of the particle anisotropy axes ($\alpha_{\mathrm{c}} = 0$). In this limit the vortex contribution vanishes, and the angular anisotropy of the spin-flip SANS cross section
reduces to
\[
I(\theta) \sim 1 - \cos(4\theta),
\]
corresponding to a pure four-fold symmetry. This state forms the lower left corner (light blue) of Fig.~\ref{fig2}.

\textit{(ii)~Strong vortex limit with aligned axes---vertical two-fold ($\mu \to \infty, \alpha_{\mathrm{c}} = 0$).}
In the limit of vanishing axial contribution ($m_0 \to 0$),
the spin-flip cross section is governed by the planar vortex term.
For a perfectly aligned orientational distribution of vortex axes
($\alpha_{\mathrm{c}} = 0$), the angular dependence reduces to
\[
I(\theta) \sim 1 - \cos(2\theta),
\]
corresponding to a vertical two-fold symmetry. This regime occupies the upper left corner (dark blue) of Fig.~\ref{fig2}.

A micromagnetic reference system illustrating this regime is discussed in Ref.~\cite{Adams2024vortex}. For moderate values of $\mu$ and finite cone angles ($\alpha_{\mathrm{c}} > 0$), for example in a remanent state, the spin-flip scattering pattern exhibits a dominant vertical two-fold symmetry with minor higher-order angular contributions.

\textit{(iii)~Strong vortex limit with isotropic axes---horizontal two-fold ($\mu \rightarrow \infty, \alpha_{\mathrm{c}} = 90^\circ$).}
For an isotropic vortex-axis distribution, we find
\[
I(\theta) \sim 3 + \cos(2\theta),
\]
producing a horizontal two-fold anisotropy. This regime occupies the upper right part of the map in Fig.~\ref{fig2} (orange).

A formally identical angular dependence was reported in dipolar-coupled nanoparticle clusters in Ref.~\cite{bender2018dipolar}. In that work, the angular anisotropy of the spin-flip SANS cross section at a small applied field of $2\; \mathrm{mT}$ followed a
\(
1 + \cos^4\theta + \sin^2\theta \cos^2\theta
= \tfrac12(3 + \cos(2\theta))
\)~anisotropy, which was interpreted in terms of antiferromagnetic-like correlations between particle moments. 

Within the present work, such an angular dependence naturally emerges from a vortex-dominated magnetization state with isotropically distributed vortex axes. While the microscopic interpretation was previously formulated in terms of dipolar correlations, the analytical structure derived here may provide an explicit magnetization-field realization reproducing the same spin-flip anisotropy.

\textit{(iv)~Strong vortex limit with ring condition---ring-like ($\mu \rightarrow \infty, 0^{\circ} \le \alpha_{\mathrm{c}} \le 90^{\circ}$).}
Between the two-fold regimes, the coefficient of $\cos(2\theta)$ vanishes when
\begin{align}
2 \cos^2\alpha_{\mathrm{c}} + 2 \cos\alpha_{\mathrm{c}} - 1 = 0,
\label{eq:RingAnisotropyRelation}
\end{align}
with the physical solution $\alpha_{\mathrm{c}} \approx 68.53^\circ$. Along this line, the angular dependence becomes isotropic,
\[
I(\theta) \sim 1,
\]
leading to a ring-like scattering pattern [Fig.~\ref{fig2} (green)]. This condition marks the analytical boundary between the two two-fold symmetry classes. Related ring-like magnetic SANS patterns have been discussed for single-vortex states in thin submicron-sized soft ferromagnetic cylinders~\cite{Metlov2016}. In contrast, the present result arises from the orientationally averaged spin-flip SANS response of a dilute vortex ensemble of spherical nanoparticles.

\begin{figure*}
    \centering
    \resizebox{0.95\textwidth}{!}{\includegraphics{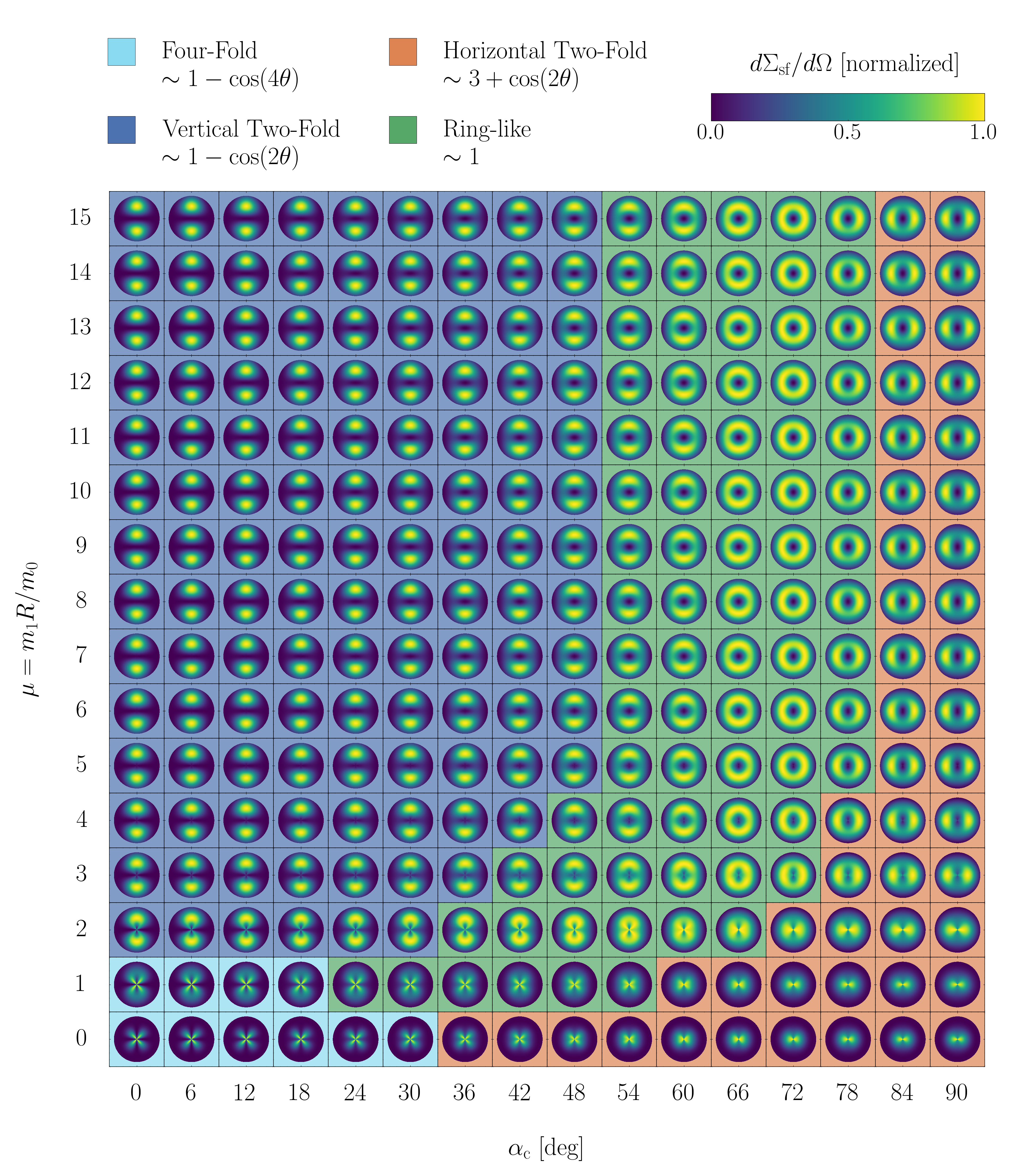}}
    \caption{
        Same as Fig.~\ref{fig2}, but based on the nonlinear hyperbolic vortex magnetization ansatz Eq.~\eqref{eq:HyperbolicVortex} and numerically computed using the \textit{NuMagSANS} software package~\cite{Adams2026NuMagSANS}. In this calculation, we used spherical nanoparticles with a radius of $R = 20 \, \mathrm{nm}$ (discretized into a $2\times2\times2 \, \mathrm{nm}^3$ grid) and $1024$~samples for the rotation angles $\alpha, \beta$. The vortex parameter $\nu$ in Eq.~\eqref{eq:HyperbolicVortex} is selected as $\nu = \mu = m_1 R/m_0$, as the hyperbolic vortex~\eqref{eq:HyperbolicVortex} reduces (in a first-order Taylor expansion at $\rho'=0$) to the linear vortex ansatz~\eqref{eq:LinearVortexAnsatz}. The resulting two-dimensional scattering cross sections are shown up to a maximum scattering vector of $q_{\mathrm{max}} = 0.25 \, \mathrm{nm}^{-1}$. The code used to generate the data is available on Zenodo~\cite{Adams2026_Landscape}.
        }
    \label{fig3}
\end{figure*}

Figure~\ref{fig2} organizes the angular solution space of Eq.~\eqref{eq:SpinFlip2DSANScrossSection} around these four analytically defined reference states. Vertical movement increases the relative vortex amplitude, whereas horizontal movement modifies the statistical distribution of vortex axes. Intermediate regions correspond to continuous interpolation between the limiting symmetries. The resulting landscape establishes a controlled link between experimentally observed two-dimensional spin-flip SANS anisotropies and specific regions in the reduced parameter space $(\mu,\alpha_{\mathrm{c}})$.

To verify that the symmetry landscape derived above is not an artifact of the assumed linear vortex ansatz, we have repeated the analysis for a nonlinear vortex profile that is motivated by micromagnetic simulations~\cite{Adams2024vortex,Adams2026minimal}. In this case, the local magnetization field is written as
\begin{align}
\mathbf{m}'(\mathbf{r}') 
&=
\operatorname{sech}(\nu \rho'/R) \, \Theta(1 - r'/R) \, \mathbf{e}_z'
\label{eq:HyperbolicVortex}
\\
&+
\tanh(\nu\rho'/R) \, \Theta(1 - r'/R) \, \mathbf{e}_{\varphi}' \nonumber .
\end{align}
Note that in \cite{Adams2024vortex,Adams2026minimal} the magnetization component along the radial direction $\mathbf{e}_{\rho}'$ is zero on the average. The parameter $\nu$ controls the width of the vortex core and relates in the small-$\rho'$ limit (first-order Taylor expansion) to the linear vortex coefficients via $m_0 = 1$ and $m_1 = \nu/R$.

In contrast to the linear ansatz, Eq.~\eqref{eq:HyperbolicVortex} does not yield a closed-form expression for the orientationally averaged spin-flip SANS cross section. We have therefore evaluated the scattering numerically by sampling the rotation angles $(\alpha,\beta)$ on the unit sphere and computing the corresponding two-dimensional cross sections using \textit{NuMagSANS}~\cite{Adams2026NuMagSANS}.

The resulting symmetry map is shown in Fig.~\ref{fig3}. Direct comparison with Fig.~\ref{fig2} demonstrates that the four symmetry classes identified for the linear model---four-fold, vertical two-fold, horizontal two-fold, and isotropic ring-like---remain similar upon replacing the linear vortex by the hyperbolic profile. The radial structure modifies the $q$~dependent weighting of the axial and vortex contributions and leads to quantitative shifts of the transition lines, but does not introduce additional angular harmonics or alters the symmetry classification.

This robustness can be understood from the Fourier structure of the scattering integrals. For a cylindrically symmetric vortex profile inside a sphere of radius $R$, the local magnetization may be written as
\begin{align}
\mathbf{m}'(\mathbf{r}')
&=
\left[
G(\rho') \, \mathbf{e}_{z}'
+
F(\rho') \, \mathbf{e}_{\varphi}'
\right] \,
\Theta(1-r'/R).
\end{align}
Its Fourier transform reads (see Appendix)
\begin{align}
\widetilde{\mathbf{m}}'(\mathbf{q}')
&=
\hat{G}(q_{\rho}', q_z') \, \mathbf{e}_{z}'
-
i \, \hat{F}(q_{\rho}', q_z') \, \mathbf{e}_{\phi}',
\\
\hat{G}(q_{\rho}', q_z')
&=
\sqrt{\frac{2}{\pi}}
\int_0^R
G(\rho') \, W(\rho',q_z') \, J_0(q_{\rho}' \rho') \, \rho' \, d\rho',
\label{eq:BesselIntegral1}
\\
\hat{F}(q_{\rho}', q_z')
&=
\sqrt{\frac{2}{\pi}}
\int_0^R
F(\rho') \, W(\rho', q_z') \, J_1(q_{\rho}' \rho') \, \rho' \, d\rho',
\label{eq:BesselIntegral2}
\\
W(\rho', q_z')
&=
\sqrt{R^2-(\rho')^2}\;
j_0\!\left( q_z' \sqrt{R^2-(\rho')^2} \right).
\end{align}
Hence, the Fourier amplitudes reduce to Bessel-weighted radial integrals: the axial component couples to $J_0$, while the azimuthal vortex component couples to $J_1$. The detailed radial profile enters only through the weights $G(\rho')$ and $F(\rho')$, whereas the finite spherical support contributes the geometric factor $W(\rho',q_z')$.

This structure shows that changing the radial profile does not generate additional symmetry-allowed angular harmonics. The azimuthal contribution, related to $\phi'$, is fixed by the cylindrical vector structure and parity of the magnetization, whereas $F$ and $G$ modify the $q'$ and $\theta'$~dependent weights in the integrals~\eqref{eq:BesselIntegral1} and \eqref{eq:BesselIntegral2} ($q_z' = q' \cos\theta', q_{\rho}' = q' \sin\theta'$). Replacing the linear vortex profile by the hyperbolic one therefore changes the radial form factors not qualitatively, as Eq.~\eqref{eq:LinearVortexAnsatz} represents the first-order Taylor expansion of Eq.~\eqref{eq:HyperbolicVortex}, but it may shift the symmetry boundaries quantitatively.

More generally, within the present spherical-support setting, the angular anisotropy classes are determined by the symmetry and vector structure of the vortex field, rather than by the detailed radial profile. This is reflected in the persistence of the topology in Fig.~\ref{fig3}.


\section{Summary and Conclusion}
\label{Summary}

We have developed a symmetry-resolved classification scheme of angular anisotropies of the two-dimensional spin-flip SANS cross section from dilute ensembles of spherical nanoparticles hosting magnetic vortex states. Based on an analytical expression for the orientationally averaged spin-flip SANS cross section derived in Ref.~\cite{Adams2024vortex}, which relies on a linear vortex ansatz and an axially symmetric distribution of vortex axes, we have analyzed the complete angular solution space in terms of the dimensionless parameters $(\mu, \alpha_{\mathrm{c}})$.

The analysis reveals that the scattering patterns organize around four analytically accessible reference states: (i)~a four-fold anisotropy corresponding to the fully saturated magnetization state, (ii)~a vertical two-fold angular symmetry arising in the limit of aligned vortex axes, (iii)~a horizontal two-fold symmetry for isotropically distributed vortex axes, and (iv)~an analytically determined ring condition at which the $\cos(2\theta)$ contribution vanishes. These limiting cases describe the full parameter space and define the symmetry landscape shown in Fig.~\ref{fig2}. Intermediate regions correspond to the continuous interpolation between these symmetry-determined regimes.

Comparison with a nonlinear (hyperbolic) vortex profile demonstrates that the identified symmetry regimes are robust with respect to the detailed radial structure of the vortex core. Although the radial profile modifies the $q$~dependent weighting of the scattering contributions and shifts quantitative boundaries, it does not introduce additional angular harmonics. The angular anisotropies are therefore governed primarily by the rotational symmetry and the statistical distribution of vortex axes, rather than by the specific functional form of the vortex profile.

The obtained symmetry landscape provides a transparent classification scheme for experimental two-dimensional spin-flip SANS data. By linking analytically controlled vortex states to experimentally accessible angular anisotropies, our work establishes a compact interpretation framework for vortex-type spin textures in nanoparticle ensembles and related dipolar-coupled systems.


\section*{Data availability}
The code and generated simulation data supporting this study are publicly available on Zenodo~\cite{Adams2026_Landscape}.




\section*{Acknowledgements}

M.P.A.\ acknowledges support by the European Union (Grant agreement ID: 101135546, MaMMoS). The views and opinions expressed are, however, those of the authors only and do not necessarily reflect those of the European Union or the European Health and Digital Executive Agency (HADEA). Neither the European Union nor the granting authority can be held responsible for them. E.M.J.\ acknowledges funding from the European Union's Horizon 2020 research and innovation program under the Marie Sk{\l}odowska-Curie actions (Grant agreement No.~101081455-YIA, HYPUSH) and from the Institute for Advanced Studies (IAS) of the University of Luxembourg.



\begin{thebibliography}{23}%
\makeatletter
\providecommand \@ifxundefined [1]{%
 \@ifx{#1\undefined}
}%
\providecommand \@ifnum [1]{%
 \ifnum #1\expandafter \@firstoftwo
 \else \expandafter \@secondoftwo
 \fi
}%
\providecommand \@ifx [1]{%
 \ifx #1\expandafter \@firstoftwo
 \else \expandafter \@secondoftwo
 \fi
}%
\providecommand \natexlab [1]{#1}%
\providecommand \enquote  [1]{``#1''}%
\providecommand \bibnamefont  [1]{#1}%
\providecommand \bibfnamefont [1]{#1}%
\providecommand \citenamefont [1]{#1}%
\providecommand \href@noop [0]{\@secondoftwo}%
\providecommand \href [0]{\begingroup \@sanitize@url \@href}%
\providecommand \@href[1]{\@@startlink{#1}\@@href}%
\providecommand \@@href[1]{\endgroup#1\@@endlink}%
\providecommand \@sanitize@url [0]{\catcode `\\12\catcode `\$12\catcode `\&12\catcode `\#12\catcode `\^12\catcode `\_12\catcode `\%12\relax}%
\providecommand \@@startlink[1]{}%
\providecommand \@@endlink[0]{}%
\providecommand \url  [0]{\begingroup\@sanitize@url \@url }%
\providecommand \@url [1]{\endgroup\@href {#1}{\urlprefix }}%
\providecommand \urlprefix  [0]{URL }%
\providecommand \Eprint [0]{\href }%
\providecommand \doibase [0]{https://doi.org/}%
\providecommand \selectlanguage [0]{\@gobble}%
\providecommand \bibinfo  [0]{\@secondoftwo}%
\providecommand \bibfield  [0]{\@secondoftwo}%
\providecommand \translation [1]{[#1]}%
\providecommand \BibitemOpen [0]{}%
\providecommand \bibitemStop [0]{}%
\providecommand \bibitemNoStop [0]{.\EOS\space}%
\providecommand \EOS [0]{\spacefactor3000\relax}%
\providecommand \BibitemShut  [1]{\csname bibitem#1\endcsname}%
\let\auto@bib@innerbib\@empty
\bibitem [{\citenamefont {Bersweiler}\ \emph {et~al.}(2019)\citenamefont {Bersweiler}, \citenamefont {Bender}, \citenamefont {Vivas}, \citenamefont {Albino}, \citenamefont {Petrecca}, \citenamefont {M\"uhlbauer}, \citenamefont {Erokhin}, \citenamefont {Berkov}, \citenamefont {Sangregorio},\ and\ \citenamefont {Michels}}]{bersweiler2019size}%
  \BibitemOpen
  \bibfield  {author} {\bibinfo {author} {\bibfnamefont {M.}~\bibnamefont {Bersweiler}}, \bibinfo {author} {\bibfnamefont {P.}~\bibnamefont {Bender}}, \bibinfo {author} {\bibfnamefont {L.~G.}\ \bibnamefont {Vivas}}, \bibinfo {author} {\bibfnamefont {M.}~\bibnamefont {Albino}}, \bibinfo {author} {\bibfnamefont {M.}~\bibnamefont {Petrecca}}, \bibinfo {author} {\bibfnamefont {S.}~\bibnamefont {M\"uhlbauer}}, \bibinfo {author} {\bibfnamefont {S.}~\bibnamefont {Erokhin}}, \bibinfo {author} {\bibfnamefont {D.}~\bibnamefont {Berkov}}, \bibinfo {author} {\bibfnamefont {C.}~\bibnamefont {Sangregorio}},\ and\ \bibinfo {author} {\bibfnamefont {A.}~\bibnamefont {Michels}},\ }\bibfield  {title} {\bibinfo {title} {Size-dependent spatial magnetization profile of manganese-zinc ferrite $\mathrm{Mn}_{0.2}\mathrm{Zn}_{0.2}\mathrm{Fe}_{2.6}\mathrm{O}_{4}$ nanoparticles},\ }\href {https://doi.org/10.1103/PhysRevB.100.144434} {\bibfield  {journal} {\bibinfo  {journal} {Phys. Rev. B}\ }\textbf {\bibinfo {volume} {100}},\ \bibinfo
  {pages} {144434} (\bibinfo {year} {2019})}\BibitemShut {NoStop}%
\bibitem [{\citenamefont {Z\'akutn\'a}\ \emph {et~al.}(2020)\citenamefont {Z\'akutn\'a}, \citenamefont {Ni$\mathrm{\check{z}}$$\mathrm{\check{n}}$ansk\'y}, \citenamefont {Barnsley}, \citenamefont {Babcock}, \citenamefont {Salhi}, \citenamefont {Feoktystov}, \citenamefont {Honecker},\ and\ \citenamefont {Disch}}]{zakutna2020}%
  \BibitemOpen
  \bibfield  {author} {\bibinfo {author} {\bibfnamefont {D.}~\bibnamefont {Z\'akutn\'a}}, \bibinfo {author} {\bibfnamefont {D.}~\bibnamefont {Ni$\mathrm{\check{z}}$$\mathrm{\check{n}}$ansk\'y}}, \bibinfo {author} {\bibfnamefont {L.~C.}\ \bibnamefont {Barnsley}}, \bibinfo {author} {\bibfnamefont {E.}~\bibnamefont {Babcock}}, \bibinfo {author} {\bibfnamefont {Z.}~\bibnamefont {Salhi}}, \bibinfo {author} {\bibfnamefont {A.}~\bibnamefont {Feoktystov}}, \bibinfo {author} {\bibfnamefont {D.}~\bibnamefont {Honecker}},\ and\ \bibinfo {author} {\bibfnamefont {S.}~\bibnamefont {Disch}},\ }\bibfield  {title} {\bibinfo {title} {{Field dependence of magnetic disorder in nanoparticles}},\ }\href {https://link.aps.org/doi/10.1103/PhysRevX.10.031019} {\bibfield  {journal} {\bibinfo  {journal} {Phys. Rev. X}\ }\textbf {\bibinfo {volume} {10}},\ \bibinfo {pages} {031019} (\bibinfo {year} {2020})}\BibitemShut {NoStop}%
\bibitem [{\citenamefont {Vivas}\ \emph {et~al.}(2020)\citenamefont {Vivas}, \citenamefont {Yanes}, \citenamefont {Berkov}, \citenamefont {Erokhin}, \citenamefont {Bersweiler}, \citenamefont {Honecker}, \citenamefont {Bender},\ and\ \citenamefont {Michels}}]{laura2020}%
  \BibitemOpen
  \bibfield  {author} {\bibinfo {author} {\bibfnamefont {L.~G.}\ \bibnamefont {Vivas}}, \bibinfo {author} {\bibfnamefont {R.}~\bibnamefont {Yanes}}, \bibinfo {author} {\bibfnamefont {D.}~\bibnamefont {Berkov}}, \bibinfo {author} {\bibfnamefont {S.}~\bibnamefont {Erokhin}}, \bibinfo {author} {\bibfnamefont {M.}~\bibnamefont {Bersweiler}}, \bibinfo {author} {\bibfnamefont {D.}~\bibnamefont {Honecker}}, \bibinfo {author} {\bibfnamefont {P.}~\bibnamefont {Bender}},\ and\ \bibinfo {author} {\bibfnamefont {A.}~\bibnamefont {Michels}},\ }\bibfield  {title} {\bibinfo {title} {{Toward understanding complex spin textures in nanoparticles by magnetic neutron scattering}},\ }\href {https://link.aps.org/doi/10.1103/PhysRevLett.125.117201} {\bibfield  {journal} {\bibinfo  {journal} {Phys. Rev. Lett.}\ }\textbf {\bibinfo {volume} {125}},\ \bibinfo {pages} {117201} (\bibinfo {year} {2020})}\BibitemShut {NoStop}%
\bibitem [{\citenamefont {Honecker}\ \emph {et~al.}(2022)\citenamefont {Honecker}, \citenamefont {Bersweiler}, \citenamefont {Erokhin}, \citenamefont {Berkov}, \citenamefont {Chesnel}, \citenamefont {Venero}, \citenamefont {Qdemat}, \citenamefont {Disch}, \citenamefont {Jochum}, \citenamefont {Michels},\ and\ \citenamefont {Bender}}]{dirkreview2022}%
  \BibitemOpen
  \bibfield  {author} {\bibinfo {author} {\bibfnamefont {D.}~\bibnamefont {Honecker}}, \bibinfo {author} {\bibfnamefont {M.}~\bibnamefont {Bersweiler}}, \bibinfo {author} {\bibfnamefont {S.}~\bibnamefont {Erokhin}}, \bibinfo {author} {\bibfnamefont {D.}~\bibnamefont {Berkov}}, \bibinfo {author} {\bibfnamefont {K.}~\bibnamefont {Chesnel}}, \bibinfo {author} {\bibfnamefont {D.~A.}\ \bibnamefont {Venero}}, \bibinfo {author} {\bibfnamefont {A.}~\bibnamefont {Qdemat}}, \bibinfo {author} {\bibfnamefont {S.}~\bibnamefont {Disch}}, \bibinfo {author} {\bibfnamefont {J.~K.}\ \bibnamefont {Jochum}}, \bibinfo {author} {\bibfnamefont {A.}~\bibnamefont {Michels}},\ and\ \bibinfo {author} {\bibfnamefont {P.}~\bibnamefont {Bender}},\ }\bibfield  {title} {\bibinfo {title} {{Using small-angle scattering to guide functional magnetic nanoparticle design}},\ }\href {http://dx.doi.org/10.1039/D1NA00482D} {\bibfield  {journal} {\bibinfo  {journal} {Nanoscale Adv.}\ }\textbf {\bibinfo {volume} {4}},\ \bibinfo {pages} {1026}
  (\bibinfo {year} {2022})}\BibitemShut {NoStop}%
\bibitem [{\citenamefont {Gerina}\ \emph {et~al.}(2023)\citenamefont {Gerina}, \citenamefont {Angotzi}, \citenamefont {Mameli}, \citenamefont {Gajdo{\v{s}}ov{\'a}}, \citenamefont {Rainer}, \citenamefont {Dopita}, \citenamefont {Steinke}, \citenamefont {Aur{\'e}lio}, \citenamefont {Vejpravov{\'a}},\ and\ \citenamefont {Z{\'a}kutn{\'a}}}]{gerina2023size}%
  \BibitemOpen
  \bibfield  {author} {\bibinfo {author} {\bibfnamefont {M.}~\bibnamefont {Gerina}}, \bibinfo {author} {\bibfnamefont {M.~S.}\ \bibnamefont {Angotzi}}, \bibinfo {author} {\bibfnamefont {V.}~\bibnamefont {Mameli}}, \bibinfo {author} {\bibfnamefont {V.}~\bibnamefont {Gajdo{\v{s}}ov{\'a}}}, \bibinfo {author} {\bibfnamefont {D.~N.}\ \bibnamefont {Rainer}}, \bibinfo {author} {\bibfnamefont {M.}~\bibnamefont {Dopita}}, \bibinfo {author} {\bibfnamefont {N.-J.}\ \bibnamefont {Steinke}}, \bibinfo {author} {\bibfnamefont {D.}~\bibnamefont {Aur{\'e}lio}}, \bibinfo {author} {\bibfnamefont {J.}~\bibnamefont {Vejpravov{\'a}}},\ and\ \bibinfo {author} {\bibfnamefont {D.}~\bibnamefont {Z{\'a}kutn{\'a}}},\ }\bibfield  {title} {\bibinfo {title} {Size dependence of the surface spin disorder and surface anisotropy constant in ferrite nanoparticles},\ }\href {http://dx.doi.org/10.1039/D3NA00266G} {\bibfield  {journal} {\bibinfo  {journal} {Nanoscale Adv.}\ }\textbf {\bibinfo {volume} {5}},\ \bibinfo {pages} {4563} (\bibinfo
  {year} {2023})}\BibitemShut {NoStop}%
\bibitem [{\citenamefont {Bersweiler}\ \emph {et~al.}(2023)\citenamefont {Bersweiler}, \citenamefont {Oba}, \citenamefont {Pratami~Sinaga}, \citenamefont {Peral}, \citenamefont {Titov}, \citenamefont {Adams}, \citenamefont {Rai}, \citenamefont {Metlov},\ and\ \citenamefont {Michels}}]{bersweilerprb2023}%
  \BibitemOpen
  \bibfield  {author} {\bibinfo {author} {\bibfnamefont {M.}~\bibnamefont {Bersweiler}}, \bibinfo {author} {\bibfnamefont {Y.}~\bibnamefont {Oba}}, \bibinfo {author} {\bibfnamefont {E.}~\bibnamefont {Pratami~Sinaga}}, \bibinfo {author} {\bibfnamefont {I.}~\bibnamefont {Peral}}, \bibinfo {author} {\bibfnamefont {I.}~\bibnamefont {Titov}}, \bibinfo {author} {\bibfnamefont {M.~P.}\ \bibnamefont {Adams}}, \bibinfo {author} {\bibfnamefont {V.}~\bibnamefont {Rai}}, \bibinfo {author} {\bibfnamefont {K.~L.}\ \bibnamefont {Metlov}},\ and\ \bibinfo {author} {\bibfnamefont {A.}~\bibnamefont {Michels}},\ }\bibfield  {title} {\bibinfo {title} {{Fingerprint of vortexlike flux closure in an isotropic Nd-Fe-B bulk magnet}},\ }\href {https://doi.org/10.1103/PhysRevB.108.094434} {\bibfield  {journal} {\bibinfo  {journal} {Phys. Rev. B}\ }\textbf {\bibinfo {volume} {108}},\ \bibinfo {pages} {094434} (\bibinfo {year} {2023})}\BibitemShut {NoStop}%
\bibitem [{\citenamefont {Titov}\ \emph {et~al.}(2025)\citenamefont {Titov}, \citenamefont {Bersweiler}, \citenamefont {Adams}, \citenamefont {Sinaga}, \citenamefont {Rai}, \citenamefont {Li\v{s}\v{c}\'{a}k}, \citenamefont {Lahr}, \citenamefont {Schmidt}, \citenamefont {Kuchkin}, \citenamefont {Haller}, \citenamefont {Suzuki}, \citenamefont {Steinke}, \citenamefont {Venero}, \citenamefont {Honecker}, \citenamefont {Kohlbrecher}, \citenamefont {Barqu\'{\i}n},\ and\ \citenamefont {Michels}}]{titov2025}%
  \BibitemOpen
  \bibfield  {author} {\bibinfo {author} {\bibfnamefont {I.}~\bibnamefont {Titov}}, \bibinfo {author} {\bibfnamefont {M.}~\bibnamefont {Bersweiler}}, \bibinfo {author} {\bibfnamefont {M.~P.}\ \bibnamefont {Adams}}, \bibinfo {author} {\bibfnamefont {E.~P.}\ \bibnamefont {Sinaga}}, \bibinfo {author} {\bibfnamefont {V.}~\bibnamefont {Rai}}, \bibinfo {author} {\bibfnamefont {\v{S}.}~\bibnamefont {Li\v{s}\v{c}\'{a}k}}, \bibinfo {author} {\bibfnamefont {M.}~\bibnamefont {Lahr}}, \bibinfo {author} {\bibfnamefont {T.~L.}\ \bibnamefont {Schmidt}}, \bibinfo {author} {\bibfnamefont {V.~M.}\ \bibnamefont {Kuchkin}}, \bibinfo {author} {\bibfnamefont {A.}~\bibnamefont {Haller}}, \bibinfo {author} {\bibfnamefont {K.}~\bibnamefont {Suzuki}}, \bibinfo {author} {\bibfnamefont {N.-J.}\ \bibnamefont {Steinke}}, \bibinfo {author} {\bibfnamefont {D.~A.}\ \bibnamefont {Venero}}, \bibinfo {author} {\bibfnamefont {D.}~\bibnamefont {Honecker}}, \bibinfo {author} {\bibfnamefont {J.}~\bibnamefont {Kohlbrecher}}, \bibinfo {author}
  {\bibfnamefont {L.~F.}\ \bibnamefont {Barqu\'{\i}n}},\ and\ \bibinfo {author} {\bibfnamefont {A.}~\bibnamefont {Michels}},\ }\bibfield  {title} {\bibinfo {title} {Spin-disorder-induced angular anisotropy in polarized magnetic neutron scattering},\ }\href {https://doi.org/10.1103/5yc2-pv4y} {\bibfield  {journal} {\bibinfo  {journal} {Phys. Rev. Lett.}\ }\textbf {\bibinfo {volume} {135}},\ \bibinfo {pages} {196706} (\bibinfo {year} {2025})}\BibitemShut {NoStop}%
\bibitem [{\citenamefont {Borchers}\ \emph {et~al.}(2025)\citenamefont {Borchers}, \citenamefont {Krycka}, \citenamefont {Bosch-Santos}, \citenamefont {de~Lima~Correa}, \citenamefont {Sharma}, \citenamefont {Carlton}, \citenamefont {Dang}, \citenamefont {Donahue}, \citenamefont {Gr\"uttner}, \citenamefont {Ivkov},\ and\ \citenamefont {Dennis}}]{borchers2025}%
  \BibitemOpen
  \bibfield  {author} {\bibinfo {author} {\bibfnamefont {J.}~\bibnamefont {Borchers}}, \bibinfo {author} {\bibfnamefont {K.}~\bibnamefont {Krycka}}, \bibinfo {author} {\bibfnamefont {B.}~\bibnamefont {Bosch-Santos}}, \bibinfo {author} {\bibfnamefont {E.}~\bibnamefont {de~Lima~Correa}}, \bibinfo {author} {\bibfnamefont {A.}~\bibnamefont {Sharma}}, \bibinfo {author} {\bibfnamefont {H.}~\bibnamefont {Carlton}}, \bibinfo {author} {\bibfnamefont {Y.}~\bibnamefont {Dang}}, \bibinfo {author} {\bibfnamefont {M.}~\bibnamefont {Donahue}}, \bibinfo {author} {\bibfnamefont {C.}~\bibnamefont {Gr\"uttner}}, \bibinfo {author} {\bibfnamefont {R.}~\bibnamefont {Ivkov}},\ and\ \bibinfo {author} {\bibfnamefont {C.~L.}\ \bibnamefont {Dennis}},\ }\bibfield  {title} {\bibinfo {title} {Magnetic anisotropy dominates over physical and magnetic structure in performance of magnetic nanoflowers},\ }\href {https://onlinelibrary.wiley.com/doi/abs/10.1002/sstr.202400410} {\bibfield  {journal} {\bibinfo  {journal} {Small Structures}\
  }\textbf {\bibinfo {volume} {6}},\ \bibinfo {pages} {2400410} (\bibinfo {year} {2025})}\BibitemShut {NoStop}%
\bibitem [{\citenamefont {R{\"o}sch}\ \emph {et~al.}(2025)\citenamefont {R{\"o}sch}, \citenamefont {Wendt}, \citenamefont {Amin}, \citenamefont {Landers}, \citenamefont {Etzkorn}, \citenamefont {Everett}, \citenamefont {M{\"u}ller-Buschbaum}, \citenamefont {Wende}, \citenamefont {Schilling}, \citenamefont {Leliaert}, \citenamefont {Honecker}, \citenamefont {Rochels}, \citenamefont {Disch},\ and\ \citenamefont {Lak}}]{disch2025}%
  \BibitemOpen
  \bibfield  {author} {\bibinfo {author} {\bibfnamefont {E.~L.}\ \bibnamefont {R{\"o}sch}}, \bibinfo {author} {\bibfnamefont {E.~L.}\ \bibnamefont {Wendt}}, \bibinfo {author} {\bibfnamefont {R.}~\bibnamefont {Amin}}, \bibinfo {author} {\bibfnamefont {J.}~\bibnamefont {Landers}}, \bibinfo {author} {\bibfnamefont {M.}~\bibnamefont {Etzkorn}}, \bibinfo {author} {\bibfnamefont {C.~R.}\ \bibnamefont {Everett}}, \bibinfo {author} {\bibfnamefont {P.}~\bibnamefont {M{\"u}ller-Buschbaum}}, \bibinfo {author} {\bibfnamefont {H.}~\bibnamefont {Wende}}, \bibinfo {author} {\bibfnamefont {M.}~\bibnamefont {Schilling}}, \bibinfo {author} {\bibfnamefont {J.}~\bibnamefont {Leliaert}}, \bibinfo {author} {\bibfnamefont {D.}~\bibnamefont {Honecker}}, \bibinfo {author} {\bibfnamefont {L.}~\bibnamefont {Rochels}}, \bibinfo {author} {\bibfnamefont {S.}~\bibnamefont {Disch}},\ and\ \bibinfo {author} {\bibfnamefont {A.}~\bibnamefont {Lak}},\ }\bibfield  {title} {\bibinfo {title} {Origin of wasp-waisted shape of magnetization
  hysteresis loops in {Co$_x$Fe$_{3-x}$O$_4$} nanoassemblies for magnetic hyperthermia},\ }\href {https://doi.org/10.1021/acsanm.5c03223} {\bibfield  {journal} {\bibinfo  {journal} {ACS Applied Nano Materials}\ }\textbf {\bibinfo {volume} {8}},\ \bibinfo {pages} {18056} (\bibinfo {year} {2025})}\BibitemShut {NoStop}%
\bibitem [{\citenamefont {Bender}\ \emph {et~al.}(2018)\citenamefont {Bender}, \citenamefont {Wetterskog}, \citenamefont {Honecker}, \citenamefont {Fock}, \citenamefont {Frandsen}, \citenamefont {Moerland}, \citenamefont {Bogart}, \citenamefont {Posth}, \citenamefont {Szczerba}, \citenamefont {Gavil{\'a}n} \emph {et~al.}}]{bender2018dipolar}%
  \BibitemOpen
  \bibfield  {author} {\bibinfo {author} {\bibfnamefont {P.}~\bibnamefont {Bender}}, \bibinfo {author} {\bibfnamefont {E.}~\bibnamefont {Wetterskog}}, \bibinfo {author} {\bibfnamefont {D.}~\bibnamefont {Honecker}}, \bibinfo {author} {\bibfnamefont {J.}~\bibnamefont {Fock}}, \bibinfo {author} {\bibfnamefont {C.}~\bibnamefont {Frandsen}}, \bibinfo {author} {\bibfnamefont {C.}~\bibnamefont {Moerland}}, \bibinfo {author} {\bibfnamefont {L.~K.}\ \bibnamefont {Bogart}}, \bibinfo {author} {\bibfnamefont {O.}~\bibnamefont {Posth}}, \bibinfo {author} {\bibfnamefont {W.}~\bibnamefont {Szczerba}}, \bibinfo {author} {\bibfnamefont {H.}~\bibnamefont {Gavil{\'a}n}}, \emph {et~al.},\ }\bibfield  {title} {\bibinfo {title} {Dipolar-coupled moment correlations in clusters of magnetic nanoparticles},\ }\href {https://doi.org/10.1103/PhysRevB.98.224420} {\bibfield  {journal} {\bibinfo  {journal} {Phys. Rev. B}\ }\textbf {\bibinfo {volume} {98}},\ \bibinfo {pages} {224420} (\bibinfo {year} {2018})}\BibitemShut {NoStop}%
\bibitem [{\citenamefont {Bender}\ \emph {et~al.}(2019)\citenamefont {Bender}, \citenamefont {Honecker},\ and\ \citenamefont {Fern{\'a}ndez~Barqu{\'i}n}}]{benderapl2019}%
  \BibitemOpen
  \bibfield  {author} {\bibinfo {author} {\bibfnamefont {P.}~\bibnamefont {Bender}}, \bibinfo {author} {\bibfnamefont {D.}~\bibnamefont {Honecker}},\ and\ \bibinfo {author} {\bibfnamefont {L.}~\bibnamefont {Fern{\'a}ndez~Barqu{\'i}n}},\ }\bibfield  {title} {\bibinfo {title} {Supraferromagnetic correlations in clusters of magnetic nanoflowers},\ }\href {https://doi.org/10.1063/1.5121234} {\bibfield  {journal} {\bibinfo  {journal} {Appl. Phys. Lett.}\ }\textbf {\bibinfo {volume} {115}},\ \bibinfo {pages} {132406} (\bibinfo {year} {2019})}\BibitemShut {NoStop}%
\bibitem [{\citenamefont {Bonilla}\ \emph {et~al.}(2017)\citenamefont {Bonilla}, \citenamefont {Lacroix},\ and\ \citenamefont {Blon}}]{Bonilla2017}%
  \BibitemOpen
  \bibfield  {author} {\bibinfo {author} {\bibfnamefont {F.~J.}\ \bibnamefont {Bonilla}}, \bibinfo {author} {\bibfnamefont {L.-M.}\ \bibnamefont {Lacroix}},\ and\ \bibinfo {author} {\bibfnamefont {T.}~\bibnamefont {Blon}},\ }\bibfield  {title} {\bibinfo {title} {Magnetic ground states in nanocuboids of cubic magnetocrystalline anisotropy},\ }\href {https://doi.org/10.1016/j.jmmm.2016.12.069} {\bibfield  {journal} {\bibinfo  {journal} {J. Magn. Magn. Mater.}\ }\textbf {\bibinfo {volume} {428}},\ \bibinfo {pages} {394} (\bibinfo {year} {2017})}\BibitemShut {NoStop}%
\bibitem [{\citenamefont {Usov}\ \emph {et~al.}(2018)\citenamefont {Usov}, \citenamefont {Nesmeyanov},\ and\ \citenamefont {Tarasov}}]{Usov2018}%
  \BibitemOpen
  \bibfield  {author} {\bibinfo {author} {\bibfnamefont {N.~A.}\ \bibnamefont {Usov}}, \bibinfo {author} {\bibfnamefont {M.~S.}\ \bibnamefont {Nesmeyanov}},\ and\ \bibinfo {author} {\bibfnamefont {V.~P.}\ \bibnamefont {Tarasov}},\ }\bibfield  {title} {\bibinfo {title} {Magnetic vortices as efficient nano heaters in magnetic nanoparticle hyperthermia},\ }\href {https://doi.org/10.1038/s41598-017-18162-8} {\bibfield  {journal} {\bibinfo  {journal} {Sci. Rep.}\ }\textbf {\bibinfo {volume} {8}},\ \bibinfo {pages} {1224} (\bibinfo {year} {2018})}\BibitemShut {NoStop}%
\bibitem [{\citenamefont {Niraula}\ \emph {et~al.}(2023)\citenamefont {Niraula}, \citenamefont {Toneto}, \citenamefont {Goya}, \citenamefont {Zoppellaro}, \citenamefont {Coaquira}, \citenamefont {Muraca}, \citenamefont {Denardin}, \citenamefont {Almeida}, \citenamefont {Knobel}, \citenamefont {Ayesh},\ and\ \citenamefont {Sharma}}]{Niraula2023_VortexNanospheres}%
  \BibitemOpen
  \bibfield  {author} {\bibinfo {author} {\bibfnamefont {G.}~\bibnamefont {Niraula}}, \bibinfo {author} {\bibfnamefont {D.}~\bibnamefont {Toneto}}, \bibinfo {author} {\bibfnamefont {G.~F.}\ \bibnamefont {Goya}}, \bibinfo {author} {\bibfnamefont {G.}~\bibnamefont {Zoppellaro}}, \bibinfo {author} {\bibfnamefont {J.~A.~H.}\ \bibnamefont {Coaquira}}, \bibinfo {author} {\bibfnamefont {D.}~\bibnamefont {Muraca}}, \bibinfo {author} {\bibfnamefont {J.~C.}\ \bibnamefont {Denardin}}, \bibinfo {author} {\bibfnamefont {T.~P.}\ \bibnamefont {Almeida}}, \bibinfo {author} {\bibfnamefont {M.}~\bibnamefont {Knobel}}, \bibinfo {author} {\bibfnamefont {A.~I.}\ \bibnamefont {Ayesh}},\ and\ \bibinfo {author} {\bibfnamefont {S.~K.}\ \bibnamefont {Sharma}},\ }\bibfield  {title} {\bibinfo {title} {Observation of magnetic vortex configuration in non-stoichiometric {Fe$_3$O$_4$} nanospheres},\ }\href {https://doi.org/10.1039/D3NA00433C} {\bibfield  {journal} {\bibinfo  {journal} {Nanoscale Adv.}\ }\textbf {\bibinfo {volume} {5}},\
  \bibinfo {pages} {5015} (\bibinfo {year} {2023})}\BibitemShut {NoStop}%
\bibitem [{\citenamefont {Moya}\ \emph {et~al.}(2024)\citenamefont {Moya}, \citenamefont {Escoda-Torroella}, \citenamefont {Rodr{\'i}guez-{\'A}lvarez}, \citenamefont {Figueroa}, \citenamefont {Garc{\'i}a}, \citenamefont {Ferrer-Vidal}, \citenamefont {Gallo-Cordova}, \citenamefont {Puerto~Morales}, \citenamefont {Aballe}, \citenamefont {Fraile~Rodr{\'i}guez}, \citenamefont {Labarta},\ and\ \citenamefont {Batlle}}]{BATLLE2024}%
  \BibitemOpen
  \bibfield  {author} {\bibinfo {author} {\bibfnamefont {C.}~\bibnamefont {Moya}}, \bibinfo {author} {\bibfnamefont {M.}~\bibnamefont {Escoda-Torroella}}, \bibinfo {author} {\bibfnamefont {J.}~\bibnamefont {Rodr{\'i}guez-{\'A}lvarez}}, \bibinfo {author} {\bibfnamefont {A.~I.}\ \bibnamefont {Figueroa}}, \bibinfo {author} {\bibfnamefont {{\'I}.}~\bibnamefont {Garc{\'i}a}}, \bibinfo {author} {\bibfnamefont {I.~B.}\ \bibnamefont {Ferrer-Vidal}}, \bibinfo {author} {\bibfnamefont {A.}~\bibnamefont {Gallo-Cordova}}, \bibinfo {author} {\bibfnamefont {M.}~\bibnamefont {Puerto~Morales}}, \bibinfo {author} {\bibfnamefont {L.}~\bibnamefont {Aballe}}, \bibinfo {author} {\bibfnamefont {A.}~\bibnamefont {Fraile~Rodr{\'i}guez}}, \bibinfo {author} {\bibfnamefont {A.}~\bibnamefont {Labarta}},\ and\ \bibinfo {author} {\bibfnamefont {X.}~\bibnamefont {Batlle}},\ }\bibfield  {title} {\bibinfo {title} {{Unveiling the crystal and magnetic texture of iron oxide nanoflowers}},\ }\href {http://dx.doi.org/10.1039/D3NR04608G} {\bibfield
   {journal} {\bibinfo  {journal} {Nanoscale}\ }\textbf {\bibinfo {volume} {16}},\ \bibinfo {pages} {1942} (\bibinfo {year} {2024})}\BibitemShut {NoStop}%
\bibitem [{\citenamefont {Jefremovas}\ \emph {et~al.}(2026)\citenamefont {Jefremovas}, \citenamefont {Calus},\ and\ \citenamefont {Leliaert}}]{jefremovas2026coercivity}%
  \BibitemOpen
  \bibfield  {author} {\bibinfo {author} {\bibfnamefont {E.~M.}\ \bibnamefont {Jefremovas}}, \bibinfo {author} {\bibfnamefont {L.}~\bibnamefont {Calus}},\ and\ \bibinfo {author} {\bibfnamefont {J.}~\bibnamefont {Leliaert}},\ }\bibfield  {title} {\bibinfo {title} {Coercivity-size map of magnetic nanoflowers: Spin disorder tunes the vortex reversal mechanism and tailors the hyperthermia sweet spot},\ }\href {https://doi.org/https://doi.org/10.1002/smsc.202500490} {\bibfield  {journal} {\bibinfo  {journal} {Small Sci.}\ }\textbf {\bibinfo {volume} {6}},\ \bibinfo {pages} {e202500490} (\bibinfo {year} {2026})}\BibitemShut {NoStop}%
\bibitem [{\citenamefont {Adams}\ and\ \citenamefont {Michels}(2026{\natexlab{a}})}]{Adams2026minimal}%
  \BibitemOpen
  \bibfield  {author} {\bibinfo {author} {\bibfnamefont {M.~P.}\ \bibnamefont {Adams}}\ and\ \bibinfo {author} {\bibfnamefont {A.}~\bibnamefont {Michels}},\ }\bibfield  {title} {\bibinfo {title} {Minimal model for vortex nucleation and reversal in spherical magnetic nanoparticles},\ }\bibfield  {journal} {\bibinfo  {journal} {arXiv preprint}\ }\href {https://doi.org/10.48550/arXiv.2601.17176} {10.48550/arXiv.2601.17176} (\bibinfo {year} {2026}{\natexlab{a}}),\ \Eprint {https://arxiv.org/abs/2601.17176} {arXiv:2601.17176 [cond-mat.mtrl-sci]} \BibitemShut {NoStop}%
\bibitem [{\citenamefont {Aharoni}(2000)}]{aharonibook}%
  \BibitemOpen
  \bibfield  {author} {\bibinfo {author} {\bibfnamefont {A.}~\bibnamefont {Aharoni}},\ }\href@noop {} {\emph {\bibinfo {title} {{Introduction to the Theory of Ferromagnetism}}}},\ \bibinfo {edition} {2nd}\ ed.\ (\bibinfo  {publisher} {Oxford University Press},\ \bibinfo {address} {Oxford},\ \bibinfo {year} {2000})\BibitemShut {NoStop}%
\bibitem [{\citenamefont {Adams}\ \emph {et~al.}(2024{\natexlab{a}})\citenamefont {Adams}, \citenamefont {Sinaga}, \citenamefont {\v{S}. Li\v{s}\v{c}\'{a}k},\ and\ \citenamefont {Michels}}]{Adams2024vortex}%
  \BibitemOpen
  \bibfield  {author} {\bibinfo {author} {\bibfnamefont {M.~P.}\ \bibnamefont {Adams}}, \bibinfo {author} {\bibfnamefont {E.~P.}\ \bibnamefont {Sinaga}}, \bibinfo {author} {\bibnamefont {\v{S}. Li\v{s}\v{c}\'{a}k}},\ and\ \bibinfo {author} {\bibfnamefont {A.}~\bibnamefont {Michels}},\ }\bibfield  {title} {\bibinfo {title} {Framework for polarized magnetic neutron scattering from nanoparticle assemblies with vortex-type spin textures},\ }\href {https://doi.org/10.1103/PhysRevB.110.014420} {\bibfield  {journal} {\bibinfo  {journal} {Phys. Rev. B}\ }\textbf {\bibinfo {volume} {110}},\ \bibinfo {pages} {014420} (\bibinfo {year} {2024}{\natexlab{a}})}\BibitemShut {NoStop}%
\bibitem [{\citenamefont {Adams}\ \emph {et~al.}(2024{\natexlab{b}})\citenamefont {Adams}, \citenamefont {Sinaga}, \citenamefont {Kachkachi},\ and\ \citenamefont {Michels}}]{adams2024signature}%
  \BibitemOpen
  \bibfield  {author} {\bibinfo {author} {\bibfnamefont {M.~P.}\ \bibnamefont {Adams}}, \bibinfo {author} {\bibfnamefont {E.~P.}\ \bibnamefont {Sinaga}}, \bibinfo {author} {\bibfnamefont {H.}~\bibnamefont {Kachkachi}},\ and\ \bibinfo {author} {\bibfnamefont {A.}~\bibnamefont {Michels}},\ }\bibfield  {title} {\bibinfo {title} {Signature of surface anisotropy in the spin-flip neutron scattering cross section of spherical nanoparticles: Atomistic simulations and analytical theory},\ }\href {https://doi.org/10.1103/PhysRevB.109.024429} {\bibfield  {journal} {\bibinfo  {journal} {Phys. Rev. B}\ }\textbf {\bibinfo {volume} {109}},\ \bibinfo {pages} {024429} (\bibinfo {year} {2024}{\natexlab{b}})}\BibitemShut {NoStop}%
\bibitem [{\citenamefont {Metlov}\ and\ \citenamefont {Michels}(2016)}]{Metlov2016}%
  \BibitemOpen
  \bibfield  {author} {\bibinfo {author} {\bibfnamefont {K.~L.}\ \bibnamefont {Metlov}}\ and\ \bibinfo {author} {\bibfnamefont {A.}~\bibnamefont {Michels}},\ }\bibfield  {title} {\bibinfo {title} {Magnetic neutron scattering by magnetic vortices in thin submicron-sized soft ferromagnetic cylinders},\ }\href {https://doi.org/10.1038/srep25055} {\bibfield  {journal} {\bibinfo  {journal} {Sci. Rep.}\ }\textbf {\bibinfo {volume} {6}},\ \bibinfo {pages} {25055} (\bibinfo {year} {2016})}\BibitemShut {NoStop}%
\bibitem [{\citenamefont {Adams}\ and\ \citenamefont {Michels}(2026{\natexlab{b}})}]{Adams2026NuMagSANS}%
  \BibitemOpen
  \bibfield  {author} {\bibinfo {author} {\bibfnamefont {M.~P.}\ \bibnamefont {Adams}}\ and\ \bibinfo {author} {\bibfnamefont {A.}~\bibnamefont {Michels}},\ }\bibfield  {title} {\bibinfo {title} {{NuMagSANS: a GPU-accelerated open-source software package for the generic computation of nuclear and magnetic small-angle neutron scattering observables of complex systems}},\ }\href {https://doi.org/10.48550/arXiv.2601.18444} {\bibfield  {journal} {\bibinfo  {journal} {J. Appl. Cryst.}\ }\textbf {\bibinfo {volume} {59}},\ \bibinfo {pages} {in press} (\bibinfo {year} {2026}{\natexlab{b}})}\BibitemShut {NoStop}%
\bibitem [{\citenamefont {Adams}(2026)}]{Adams2026_Landscape}%
  \BibitemOpen
  \bibfield  {author} {\bibinfo {author} {\bibfnamefont {M.~P.}\ \bibnamefont {Adams}},\ }\bibfield  {title} {\bibinfo {title} {{Angular anisotropy landscape of vortex ensembles in polarized small-angle neutron scattering --- Dataset}},\ }\href {https://doi.org/10.5281/zenodo.19348079} {10.5281/zenodo.19348079} (\bibinfo {year} {2026})\BibitemShut {NoStop}%
\end{thebibliography}

%

\appendix*

\section{Fourier representation of cylindrically symmetric vortex profiles}
\label{sec:app_vortex_fourier}

To clarify why the symmetry regimes remain robust under changes of the radial vortex-core profile, we summarize the Fourier structure of a cylindrically symmetric vortex texture inside a spherical particle of radius $R$. For notational simplicity, we use nonprimed coordinates throughout this appendix.

We consider a magnetization field of the form
\begin{align}
\mathbf{m}(\mathbf{r})
=
\left[
G(\rho) \, \mathbf{e}_z
+
F(\rho) \, \mathbf{e}_{\varphi}
\right]
\, \Theta(1 - r/R),
\label{eq:app_vortex_m_real}
\end{align}
where
\begin{align}
\rho &= r \sin\vartheta, &
z &= r \cos\vartheta,&
r^2 &= \rho^2 + z^2,
\end{align}
and $\Theta$ denotes the Heaviside step function. In cylindrical coordinates, the unit vectors take on the following form
\begin{align}
\mathbf{e}_z
&=
\{
0, \,
0, \,
1
\},
\\
\mathbf{e}_{\varphi}
&=
\{
-\sin\varphi, \,
\cos\varphi, \,
0
\}.
\end{align}
The Fourier transform is defined by
\begin{align}
\widetilde{\mathbf{m}}(\mathbf{q})
=
\frac{1}{(2\pi)^{3/2}}
\int_{\mathbb{R}^3}
\mathbf{m}(\mathbf{r})
\mathrm{e}^{-i\mathbf{q} \cdot \mathbf{r}} \, d^3r.
\label{eq:app_vortex_ft_def}
\end{align}
Writing
\begin{align}
\mathbf{q} &= \{ q_{\rho} \cos\phi, \, q_{\rho} \sin\phi, \, q_z \},
\\
\mathbf{r} &= \{ \rho \cos\varphi , \, \rho \sin\varphi, \, z \}
\end{align}
one has
\begin{align}
\mathbf{q} \cdot \mathbf{r}
=
q_{\rho} \rho \cos(\varphi-\phi) + q_z z.
\end{align}
Since the support is restricted to the sphere $r<R$, the Fourier integral becomes
\begin{align}
\widetilde{\mathbf{m}}(\mathbf{q})
&=
\frac{1}{(2\pi)^{3/2}}
\int_0^{2\pi}
\int_0^R
\int_{-\sqrt{R^2-\rho^2}}^{\sqrt{R^2-\rho^2}}
\begin{bmatrix}
-F(\rho) \sin\varphi\\
\phantom{-} F(\rho) \cos\varphi\\
G(\rho)
\end{bmatrix}
\nonumber\\
&\qquad\times
\mathrm{e}^{-iq_{\rho}\rho\cos(\varphi-\phi)}
\mathrm{e}^{-iq_z z}\,
dz \, \rho \, d\rho \, d\varphi.
\label{eq:app_vortex_ft_cyl}
\end{align}
The $z$-integration yields
\begin{align}
\int_{-\sqrt{R^2-\rho^2}}^{\sqrt{R^2-\rho^2}}
\mathrm{e}^{-iq_z z} \, dz
=
2\sqrt{R^2-\rho^2} \,
j_0\!\left( q_z \sqrt{R^2-\rho^2} \right),
\label{eq:app_vortex_zint}
\end{align}
where $j_0(x) = \sin(x)/x$ is the spherical Bessel function of order zero. The angular integrals are
\begin{align}
\int_0^{2\pi}
\mathrm{e}^{-iq_{\rho} \rho \cos(\varphi-\phi)} \, d\varphi
&=
2\pi J_0(q_{\rho} \rho),
\\
\int_0^{2\pi}
\cos\varphi \,
\mathrm{e}^{-iq_{\rho} \rho \cos(\varphi-\phi)} \, d\varphi
&=
- \, i \, 2\pi \cos\phi \, J_1(q_{\rho} \rho),
\\
\int_0^{2\pi}
\sin\varphi \,
\mathrm{e}^{-iq_{\rho} \rho \cos(\varphi-\phi)} \, d\varphi
&=
- \, i \, 2\pi \sin\phi \, J_1(q_{\rho} \rho),
\end{align}
where the $J_n$ represent the cylindrical Bessel functions.

Collecting terms, one obtains
\begin{align}
\widetilde{\mathbf{m}}(\mathbf{q})
=
\hat{G}(q_{\rho}, q_z) \, \mathbf{e}_z
-
i \, \hat{F}(q_{\rho}, q_z) \, \mathbf{e}_{\phi},
\label{eq:app_vortex_ft_compact}
\end{align}
where
\begin{align}
\mathbf{e}_{\phi}
=
\{
-\sin\phi, \,
\cos\phi, \,
0
\},
\end{align}
and
\begin{align}
\hat{G}(q_{\rho}, q_z)
&=
\sqrt{\frac{2}{\pi}}
\int_0^R
G(\rho) \,
W(\rho,q_z) \,
J_0(q_{\rho} \rho) \,
\rho \, d\rho,
\label{eq:app_vortex_Ghat}
\\
\hat{F}(q_{\rho}, q_z)
&=
\sqrt{\frac{2}{\pi}}
\int_0^R
F(\rho) \,
W(\rho, q_z) \,
J_1(q_{\rho} \rho) \,
\rho \, d\rho,
\label{eq:app_vortex_Fhat}
\\
W(\rho,q_z)
&=
\sqrt{R^2-\rho^2}\,
j_0\!\left( q_z \sqrt{R^2-\rho^2} \right).
\label{eq:app_vortex_W}
\end{align}
Equations~\eqref{eq:app_vortex_ft_compact} $-$ \eqref{eq:app_vortex_W} show that the axial part $G(\rho)$ of the vortex profile couples to $J_0$, while the azimuthal part $F(\rho)$ couples to $J_1$. The radial profile functions $G(\rho)$ and $F(\rho)$ therefore affect only the weights of the integrals, whereas the spherical support enters through the function $W(\rho,q_z)$.

\end{document}